\documentstyle[twocolumn,prl,aps]{revtex}

\tighten
\begin{document}
\draft

\title{General-covariant constraint-free evolution system
for Numerical Relativity}
\author{C.~Bona, T.~Ledvinka$^{\dag}$, C.~Palenzuela
       and M.~\v Z\' a\v cek$^{\dag}$}

\address{
Departament de Fisica, Universitat de les Illes Balears,
Ctra de Valldemossa km 7.5, 07071 Palma de Mallorca, Spain\\
$^{\dag}$Institute of Theoretical Physics, Faculty of Mathematics
and Physics, Charles University, V Hole\v{s}ovi\v{c}k\'ach 2, 180
00 Prague 8, Czech Republic}

\maketitle

\begin{abstract}
A general covariant extension of Einstein's field equations is
considered with a view to Numerical Relativity applications. The
basic variables are taken to be the metric tensor and an
additional four-vector. The extended field equations, when
supplemented by suitable coordinate conditions, determine the time
evolution of all these variables without any constraint.
Einstein's solutions are recovered when the additional four-vector
vanishes, so that the energy and momentum constraints hold true.
The extended system is well posed when using the natural extension
of either harmonic coordinates or the harmonic slicing condition
in normal coordinates

\end{abstract}

\section{Introduction}

Einstein's field equations can be understood as a set of ten
second order partial differential equations on the ten unknown
metric coefficients $g_{\mu \nu}$
\begin{equation}\label{Einstein}
  R_{\mu \nu} = 8\; \pi\; (T_{\mu \nu} - \frac{1}{2} T\; g_{\mu \nu})
\end{equation}
But in General Relativity, like in other field theories, physical
solutions are given instead by an equivalence class of
mathematical expressions which are related one to another by gauge
transformations. The peculiarity of General Relativity is that the
gauge group  is that of the general (smooth) coordinate
transformations
\begin{equation}\label{4covariance}
  y^{\mu} = f^{\mu}(x^{\nu})
\end{equation}
This allows one to use the coordinate gauge freedom
(\ref{4covariance}) to fix up to four of the ten metric
coefficients (kinematical degrees of freedom) so that only the
remaining ones (dynamical degrees of freedom) can be actually
determined by the field equations (\ref{Einstein}). It follows
that four of these equations just provide additional constraints
on the dynamical degrees of freedom: they can be understood as
first integrals related with energy and momentum conservation.

This complex structure becomes more transparent in the context of
the 3+1 formalism \cite{Ch56,ADM62}, where one considers the
space-time sliced by $t=constant$ hypersurfaces. The line element
can then be written as
\begin{eqnarray}\label{line4D}
    ds^2 &=& - \alpha^2\;dt^2  \nonumber \\
    &+& \gamma_{ij}\;(dx^i+\beta^i\;dt)\;(dx^j+\beta^j\;dt)
     \;\;\;\;i,j=1,2,3
\end{eqnarray}
where the lapse $\alpha$ and the shift $\beta^i$ are the
kinematical degrees of freedom, whereas the coefficients
$\gamma_{ij}$ of the metric induced on the $t=constant$
hypersurfaces are the dynamical ones. One can use the extrinsic
curvature (second fundamental form) of these slices
\begin{equation}\label{K}
  K_{ij}\equiv - \frac{1}{2\;\alpha}\;
  (\partial_t -{\cal L}_{\beta})\; \gamma_{ij}
\end{equation}
to decompose the field equations (\ref{Einstein}) into six
evolution equations
\begin{eqnarray}\label{evolve_K}
   (\partial_t -{\cal L}_{\beta}) K_{ij} &=& -\nabla_i\alpha_j \\
    &+& \alpha\;
    [{}^{(3)}R_{ij}-2K^2_{ij}+trK\;K_{ij}]
\nonumber
\end{eqnarray}
(ADM evolution equations) and four constraints
\begin{eqnarray}
  ^{(3)}R - tr(K^2) + (trK)^2 = 0
\label{energy_constraint}  \\
  \nabla_k\;({K^k}_{i}-trK\;{\delta^k}_i) = 0
\label{momentum_constraint}
\end{eqnarray}
where we have restricted ourselves to the vacuum case for
simplicity.

The 3+1 formalism (\ref{line4D}-\ref{momentum_constraint}) is
specially suited for Numerical Relativity applications. In this
context, one usually takes advantage of the fact that the energy
and momentum constraints
(\ref{energy_constraint},\ref{momentum_constraint}) are first
integrals of (\ref{K},\ref{evolve_K}). This allows one to impose
(\ref{energy_constraint},\ref{momentum_constraint}) on the initial
data only ('free evolution' approach), but not during time
evolution. One is dealing then with an extended set of solutions.
The resulting simulations will represent true Einstein's solutions
only to the extent to which the constraints
(\ref{energy_constraint},\ref{momentum_constraint}) are actually
preserved during numerical evolution.

This approach raises some concerns (see Ref. \cite{LRReula}).
First of all, the set of 'extended solutions' is not determined in
a general covariant way: equations (\ref{K},\ref{evolve_K}) just
translate the space components of the original field equations
(\ref{Einstein}). And the space components of four-dimensional
tensors are covariant only under the restricted subgroup of
coordinate transformations that preserve the time slicing (3+1
covariance):
\begin{equation}\label{3covariance}
    t'= h(t) ,\;\; y^{i} = f^i(x^j,t)
\end{equation}
Moreover, even in a given coordinate system, there are many ways
of extending the space of solutions, because the set of six
evolution equations (\ref{evolve_K}) is not univocally defined in
that context: one can actually combine either the energy
constraint (\ref{energy_constraint}) or the momentum constraint
(\ref{momentum_constraint}) with the original evolution equations
(\ref{evolve_K}) to get a different set of evolution equations.
This fact has been recently used by Kidder, Scheel and Teukolsky
(KST) to obtain a multiparameter family of evolution systems
\cite{KST01}.

This ambiguity was previously used in a different way by Bona and
Massó \cite{BM92,BM95}, and also by Shibata and Nakamura
\cite{SN95} and Baumgarte and Shapiro \cite{BS99} (BSSN system).
The key idea in these works was to introduce three supplementary
dynamical variables whose evolution equations were obtained by
using the momentum constraint (\ref{momentum_constraint}). As far
as these works were focused on Numerical Relativity applications,
the supplementary quantities were introduced in an 'ad hoc' way,
breaking the 3+1 covariance of the formalism. Only very recently
\cite{BLP02} the same idea has been implemented in a 3+1 covariant
way: the extra quantities are given by a three-dimensional 'zero'
vector $Z_i$ which vanishes for Einstein's solutions. During
numerical evolution, however, non-zero values of $Z_i$ arise due
to truncation errors so that the resulting numerical codes
actually deal with an extended set of solutions.

The next logical step along this line would be to introduce one
more supplementary quantity, let us call it $\Theta$, whose
evolution equation would be obtained by using the energy
constraint (\ref{energy_constraint}) in the same way as momentum
constraint was used in \cite{BLP02} to evolve $Z_i$. The new
quantity could be then coupled to the other equations in many
different ways so that many more new parameters will appear in the
resulting evolution system (see for instance Ref. \cite{FR99} for
a similar approach).

We will not go that way in the present work. We prefer to preserve
general covariance in our extension of Einstein's field equations.
Instead of a separate three-vector $Z_i$ and three-scalar
$\Theta$, we shall consider a covariant 'zero' four-vector
$Z_{\mu}$ such that their space components coincide with $Z_i$
and
\begin{equation}\label{theta}
    \Theta \equiv n_{\mu}\; Z^{\mu} = \alpha\; Z^0
\end{equation}
where $n_{\mu}$ is the unit normal to the $t=constant$ slices. The
purpose of this work is to use $Z_{\mu}$ to extend in a covariant
way Einstein's equations so that the extended system contains only
evolution equations, without any constraint. Einstein's solutions
will be recovered for $Z_{\mu} = 0$. The four-vector $Z_{\mu}$
will then provide a simple way to measure of the quality of
numerical simulations: energy constraint violations can be
monitored with $\Theta^2$, whereas one can use the norm of $Z_i$
to deal with momentum constraint deviations. These quantities can
be combined to form a four-dimensional scalar:
\begin{equation}\label{Z2}
    - Z_{\mu} Z^{\mu} = \Theta^2 - \gamma^{ij} Z_i Z_j
\end{equation}

\section{Extending the field equations}
We propose to extend the field equations (\ref{Einstein}) in the
following way
\begin{equation}\label{Einstein_extended}
  R_{\mu \nu} + \nabla_{\mu} Z_{\nu} + \nabla_{\nu} Z_{\mu} =
  8\; \pi\; (T_{\mu \nu} - \frac{1}{2} T\; g_{\mu \nu})
\end{equation}
where we have rescaled the four-vector $Z_{\mu}$  to clean
(\ref{Einstein_extended}) from any arbitrary parameter. The Cauchy
problem of the extended equations (\ref{Einstein_extended}) is
simpler than the corresponding one for (\ref{Einstein}). The space
components of (\ref{Einstein_extended}) provide again second order
evolution equations for the dynamical degrees of freedom in the
metric. In the 3+1 language, one has
\begin{eqnarray}\label{evolve_K_extended}
   (\partial_t - {\cal L}_{\beta}) K_{ij} &=& -\nabla_i\alpha_j
   + \alpha\; [{}^{(3)}R_{ij} + \nabla_i Z_j + \nabla_j Z_i \\
   &-& 2\;K^2_{ij}+ (trK - 2\; \Theta)\;K_{ij}]
\nonumber
\end{eqnarray}
But now the remaining components can be combined with
(\ref{evolve_K_extended}) to provide, instead of constraints,
first order evolution equations for the components of $Z_{\mu}$,
namely
\begin{eqnarray}
 (\partial_t -{\cal L}_{\beta}) \Theta &=& \frac{\alpha}{2}\;
 [{}^{(3)}R + (trK - 2\; \Theta)\;trK
\label{evolve_theta}  \\
 &-& tr(K^2) + 2\; \nabla_k Z^k  - 2\; Z^k {\alpha}_k/\alpha]
\nonumber  \\
 (\partial_t -{\cal L}_{\beta}) Z_i &=& \alpha\; [\nabla_j\;({K_i}^j
  -{\delta_i}^j trK) + \partial_i \Theta
\label{evolve_Z} \\
   &-& 2\; {K_i}^j\; Z_j  - \Theta\; {\alpha}_i/\alpha]
\nonumber
\end{eqnarray}
where here again we have restricted ourselves to the vacuum case.

The contracted Bianchi identities, when applied to
eq.(\ref{Einstein_extended}), lead to a simple equation for
$Z_{\mu}$
\begin{equation}\label{laplacian}
   {}^L \Box Z_{\mu} \equiv g^{\rho \sigma} \nabla_{\rho}
   \nabla_{\sigma} Z_{\mu} + R_{\mu \nu}\; Z^{\nu} = 0
\end{equation}
where ${}^L \Box$ stands for the four-dimensional Lichnerowicz
Laplacian. This equation, of course, can be also derived from
(\ref{evolve_K_extended}-\ref{evolve_Z}), but the converse is not
true: the first order evolution equations
(\ref{evolve_theta},\ref{evolve_Z}) for $Z_{\mu}$ can not be
derived from (\ref{evolve_K_extended}) and the second order
equation (\ref{laplacian}).

\section{Coordinate conditions}
In order to analyze the causal structure of the extended equations
(\ref{Einstein_extended}), we will use de DeDonder \cite{D21}
expression of the Ricci tensor to write down the principal part,
namely
\begin{equation}\label{alembert}
 -\Box\; g_{\mu \nu} + \partial_{\mu} (\Gamma_{\nu} + 2\; Z_{\nu})
   + \partial_{\nu} (\Gamma_{\mu} + 2\; Z_{\mu}) = ...
\end{equation}
where here the box symbol stands for the d'Alembert operator on
functions and we have noted as usual $\Gamma^{\mu} \equiv g^{\rho
\sigma} {\Gamma^{\mu}}_{\rho \sigma}$. Comparing with the wave
equation for $g_{\mu \nu}$, it is clear that we can obtain a well
posed system if we take out the additional terms in
(\ref{alembert}) by using the following extension of the well
known harmonic coordinate conditions:
\begin{equation}\label{harmonic_x0_extended}
 \Box\; x^{\mu} = - \Gamma^{\mu} = 2\; Z^{\mu}
\end{equation}
so that the four-vector $Z_{\mu}$ can be interpreted in this
context as providing 'gauge sources', along the lines sketched in
ref \cite{LRFriedrich}.

Harmonic coordinates, including their extension
(\ref{harmonic_x0_extended}), are not flexible enough to be used
in most Numerical Relativity applications. A more suitable choice
is the 'harmonic slicing', in which the time coordinate is again
assumed to be an harmonic function, but the space coordinates are
chosen so that the mixed components $g_{0i}$ vanish. We propose
here to keep in the extended case the time component of
(\ref{harmonic_x0_extended}), that is
\begin{equation}\label{harmonic_slicing}
 \Box\; x^{0} = - \Gamma^{0} = 2\; Z^{0}, \;\; g_{0i}=0
\end{equation}
which can be translated into the 3+1 language as
\begin{equation}\label{evolve_alpha}
 (\partial_t -{\cal L}_{\beta}) \ln \alpha =
 - \alpha\; (trK-2\; \Theta),  \;\; \beta^i=0
\end{equation}
The mixed time-space components of (\ref{alembert}) read now
\begin{equation}\label{evolve_mixed}
 \partial_0 (\Gamma_i + 2\; Z_i) = ...
\end{equation}
This means that in the harmonic slicing case
(\ref{harmonic_slicing}) there are three degrees of freedom,
directly related with the space components of $Z$, with
characteristic lines orthogonal to the $t=constant$ slices. This
'standing' modes can be decoupled from the remaining ones by
taking an extra time derivative of (\ref{alembert}). This means
that the resulting system will be of third order in $\alpha$ and
$g_{ij}$ (second order in $\Theta$ and $K_{ij}$). Allowing for
(\ref{alembert},\ref{evolve_alpha}), one gets after some algebra
\begin{equation}\label{harmonic_light}
 \Box\; \Theta = ... \;\;\;\;\:\:\: \Box K_{ij}=...
\end{equation}
so that the corresponding characteristic lines are on the light
cones. It follows from (\ref{evolve_mixed},\ref{harmonic_light})
that the resulting system (with that extra time derivative) is
well posed (see Ref. \cite{ChR83} for a similar development, in the
3+1 language, from Einstein's equations (\ref{Einstein})).

Although numerical simulations are beyond the scope of the present
letter, we will comment here some extensions of this work which
are specially suited for Numerical Relativity. The first one is
that harmonic slicing is often generalized in that context by
including an extra factor $f(\alpha)$ in the right-hand-side of
equation (\ref{evolve_alpha}). The effect of this minor change is
twofold: from the theoretical point of view, the gauge-related
degrees of freedom no longer propagate with light speed (gauge
speed) so that the first equation in (\ref{harmonic_light}) no
longer holds true; from the numerical point of view, higher gauge
speed increases the singularity avoidance properties of the gauge
\cite{1+log}.

The second extension is to consider the first space derivatives
\begin{equation}\label{A_D}
    A_k \equiv \partial_k (\ln \alpha), \;\;
    D_{kij} \equiv \frac{1}{2}\; \partial_k \gamma_{ij}
\end{equation}
as independent dynamical variables with evolution equations
\begin{eqnarray}
   &\partial_t& A_k + \partial_k[\alpha\;f(trK-2\;\Theta)]= 0
\label{evolve_A} \\
   &\partial_t& D_{kij} + \partial_k[\alpha\; K_{ij}] =0
\label{evolve_D}
\end{eqnarray}
so that the set
(\ref{evolve_K_extended}-\ref{evolve_Z},\ref{evolve_A},\ref{evolve_D})
forms a first order hyperbolic evolution system, which is
obviously much more adapted to Numerical Relativity applications
than the above mentioned third order system. In addition, the
causal structure of first order systems is much easier to analyze
\cite{LRReula} and one can take advantage of the characteristic
field decomposition to devise stable numerical algorithms.\\

{\em Acknowledgements: This work has been supported by the EU Programme
'Improving the Human Research Potential and the Socio-Economic
Knowledge Base' (Research Training Network Contract (HPRN-CT-2000-00137),
by the Spanish Ministerio de Ciencia y Tecnologia through the research
grant number BFM2001-0988 and by a grant from the Conselleria d'Innovacio
i Energia of the Govern de les Illes Balears.}

\bibliographystyle{prsty}

\begin{thebibliography}{99}

\bibitem{Ch56} Y.~Choquet-Bruhat, J.~Rat.~Mee.~Analysis {\bf 5},
         951 (1956).

\bibitem{ADM62} R.~Arnowit, S.~Deser and C.~W.~Misner, {\em Gravitation:
        an introduction to current research}, ed. L.~Witten,
        Wiley, New York (1962).

\bibitem{LRReula} O.~Reula, {\em Hyperbolic Methods for Einstein's
        Equations}, Living Reviews in Relativity.

\bibitem{KST01} L. E. Kidder, M. A. Scheel and
         S. A. Teukolsky, Phys. Rev. D {\bf 64}, 064017 (2001).

\bibitem{BM92} C.~Bona and J.~Mass\'o,
        Phys.~Rev.~Lett. {\bf 68} 1097 (1992)

\bibitem{BM95} C. Bona, J. Mass\'o, E. Seidel
        and J. Stela, Phys. Rev. Lett. {\bf 75} 600 (1995).

\bibitem{SN95} M.~Shibata and T.~Nakamura,
        Phys.~Rev.~D {\bf 52} 5428 (1995).

\bibitem{BS99} T.~W.~Baumgarte and S.~L.~Shapiro,
        Phys.~Rev.~D {\bf 59} 024007 (1999).

\bibitem{BLP02} C.~Bona, T.~Ledvinka and C.~Palenzuela,
        gr-qc/0208087 (accepted at Phys.~Rev.~D).

\bibitem{FR99} S.~Frittelli and O.~A.~Reula,
        J.~Math.~Phys. {\bf 40} 5143 (1999).

\bibitem{D21} T.~De.~Donder, {\em La Gravifique Einstenienne}, 
       Gauthier-Villars, Paris (1921).\\
       V.~A.~Fock, {\em Theory of space, time and gravitation},
       Pergamon, London (1959).

\bibitem{LRFriedrich} H.~Friedrich and A.~Rendall {\em The
       Cauchy Problem for the Einstein Equations},
       Living Reviews in Relativity.

\bibitem{ChR83} Y. Choquet-Bruhat, Y. Ruggeri,
        Commun. Math. Physics {\bf 89}, 269 (1983).

\bibitem{1+log} A.~Abrahams et al, Phys.~Rev.~D{\bf 45}, 3544 (1992).\\
        P.~Anninos et al, Phys.~Rev.~Lett. {\bf 71} 2851 (1993).

\end{thebibliography}

\end{document}